\pdfoutput=1
\documentclass[preprint,showpacs,preprintnumbers,amsmath,amssymb,floatfix,endfloats*]{revtex4}

\usepackage{graphicx}
\usepackage{dcolumn}
\usepackage{bm}
\usepackage{amsfonts}
\usepackage{hyperref}
\usepackage{cleveref}

\DeclareMathAlphabet{\mathsfsl}{OT1}{cmr}{bx}{it}
\begin{document}
\title{Molecular dynamics simulations of the rotational and
translational diffusion of a Janus rod-shaped nanoparticle}
\author{Ali Kharazmi$^{1}$ and Nikolai V. Priezjev$^{2}$}
\affiliation{$^{1}$Department of Mechanical Engineering, Michigan
State University, East Lansing, Michigan 48824}
\affiliation{$^{2}$Department of Mechanical and Materials
Engineering, Wright State University, Dayton, Ohio 45435}
\date{\today}
%
\begin{abstract}

The diffusion of a Janus rod-shaped nanoparticle in a dense
Lennard-Jones fluid is studied using molecular dynamics (MD)
simulations.  The Janus particle is modeled as a rigid cylinder
whose atoms on each half-side have different interaction energies
with fluid molecules, thus comprising wetting and nonwetting
surfaces.  We found that both rotational and translational diffusion
coefficients are larger for Janus particles with higher wettability
contrast, and these values are bound between the two limiting cases
of uniformly wetting and nonwetting particles.   It was also shown
that values of the diffusion coefficients for displacements parallel
and perpendicular to the major axis of a uniformly wetting particle
agree well with analytical predictions despite a finite slip at the
particle surface present in MD simulations. It was further
demonstrated that diffusion of Janus particles is markedly different
from that of uniform particles; namely, Janus particles
preferentially rotate and orient their nonwetting sides along the
displacement vector to reduce drag. This correlation between
translation and rotation is consistent with the previous results on
diffusive dynamics of a spherical Janus particle with two
hemispheres of different wettability.

\end{abstract}

\pacs{68.08.-p, 66.20.-d, 83.10.Rs}


\maketitle

\section{Introduction}

The design of functional nanomaterials, with diverse applications in
biomedicine, optoelectronics and microfiltration, requires a
detailed understanding of the diffusion process of anisotropic
particles during their self-assembly in bulk fluid and at
interfaces~\cite{Busseron13}.  The synthesis of Janus nanoparticles
often involves a masking step where particles are temporarily
trapped at the interface between two phases and only one side can be
treated chemically leading to asymmetric
functionalization~\cite{Lattuada11}.  It was shown numerically that
structural evolution of polymer blends can be significantly
influenced by orientation of Janus nanorods relative to the phase
interface, and, thus, it allows fabrication of polymer
nanocomposites with robust photovoltaic and mechanical
properties~\cite{Xu12,Li13}.  Common mechanisms of colloidal
self-assembly include depletion-assisted structure formation, where
effective forces between neighboring colloidal particles arise due
to decreasing volume available to the depletant, and
shape-complementary colloidal suspensions where only particles with
matching building blocks bind together~\cite{Pine13}. Interestingly,
a variety of prescribed crystal or gel phases can be obtained via
`programmable' self-assembly of DNA-grafted particles due to
formation of bridges between neighboring particles~\cite{Rogers16}.
Regardless of the specific interaction between particles, a complete
picture of the diffusive motion even of isolated anisotropic
particles is still missing.

\vskip 0.05in

The results of equilibrium molecular dynamics simulations have shown
that the diffusion dynamics of a single spherical particle depends
on its wetting properties, local boundary conditions, mass and size
relative to the solvent molecules, as well as viscosity and
temperature of the
solvent~\cite{Levesque00,Schmidt03,Kapral04,Schmidt04,Levesque07,Li09,Shin10,Chakraborty11,Ohtori16}.
Originally, it was discovered by Alder and Wainwright~\cite{Alder70}
that the velocity autocorrelation function exhibits a characteristic
power-law decay at long times due to hydrodynamic coupling to the
solvent.    More recently, it was demonstrated that in the presence
of a liquid-solid interface, the power-law exponent is changed from
the bulk value $-3/2$ to $-5/2$, while the amplitude of the velocity
autocorrelation function increases for diffusive motion near
slipping boundary~\cite{Izabela15,Felderhof05}.  It was also shown
that as the particle size decreases down to a few molecular
diameters, the Stokes-Einstein relation breaks down, and the
effective radius of the particle might increase due to formation of
an adjacent fluid layer if the surface energy is sufficiently
large~\cite{Li09}. Despite significant computational efforts, the
exact relationship between the local slip at confining boundaries
and the position-dependent diffusion coefficient has yet to be
established.

\vskip 0.05in

In recent years, various aspects of diffusive motion of a single
Janus particle in the bulk and at liquid interfaces were studied
using continuum~\cite{Swan08,Willmott08,Willmott09,Chan13} and
molecular dynamics~\cite{Drazer15,Kharazmi15,Zadeh16,Archereau16}
simulations.  In addition, a high-speed experimental technique to
track translation and rotation of colloidal particles in three
dimensions was recently developed and used to determine accurately
the diffusion coefficients of silica rods and spherical Janus
particles in water~\cite{Manoharan14}.    Depending on the geometry
of Janus particles (spherical, rod- or disk-like) and the degree of
amphiphilicity, it was observed that particle rotational dynamics at
sheared liquid-liquid interfaces involves either a smooth tilt or a
tumbling motion~\cite{Zadeh16}.    In the previous study, the
translational and rotational diffusion of a spherical Janus particle
in a dense fluid was investigated using molecular dynamics
simulations~\cite{Kharazmi15}.   In particular, it was shown that
Janus particles with lower surface energy diffuse faster, and their
nonwetting hemispheres tend to orient along the displacement vector
of the center of mass during the rotational relaxation
time~\cite{Kharazmi15}.   However, the combined effect of particle
shape anisotropy and wettability contrast on diffusion remains not
fully understood.

\vskip 0.05in

In this paper, we investigate the diffusion of a rod-shaped Janus
particle in the limit of infinite dilution using molecular dynamics
simulations.    The particle consists of atoms rigidly fixed at the
lattice sites that form a cylinder, which undergoes a diffusive
motion under random forces from fluid molecules.   In our model, the
wall-fluid interaction energy at each half-side can be adjusted to
control local wetting properties at the particle surface.  We show
that with increasing wettability contrast, the translational and
angular displacements become larger and the effective center of
rotation is displaced toward the wetting end.   We also demonstrate
that a Janus particle on average is rotated by its nonwetting side
along the displacement vector of the center of mass to reduce drag.

\vskip 0.05in

The paper is organized as follows.   The details of the simulation
procedure and the particle model are described in the next section.
In Sec.\,\ref{sec:Results}, we report the fluid density profiles
around particles, determine the local slip boundary conditions, and
estimate translational and rotational diffusion coefficients from
the particle trajectories and make a comparison with theoretical
predictions. Brief conclusions are given in the last section.

\section{Simulation Method}
\label{sec:MD_Model}

\begin{figure}[t]
\includegraphics[width=10.cm,angle=0]{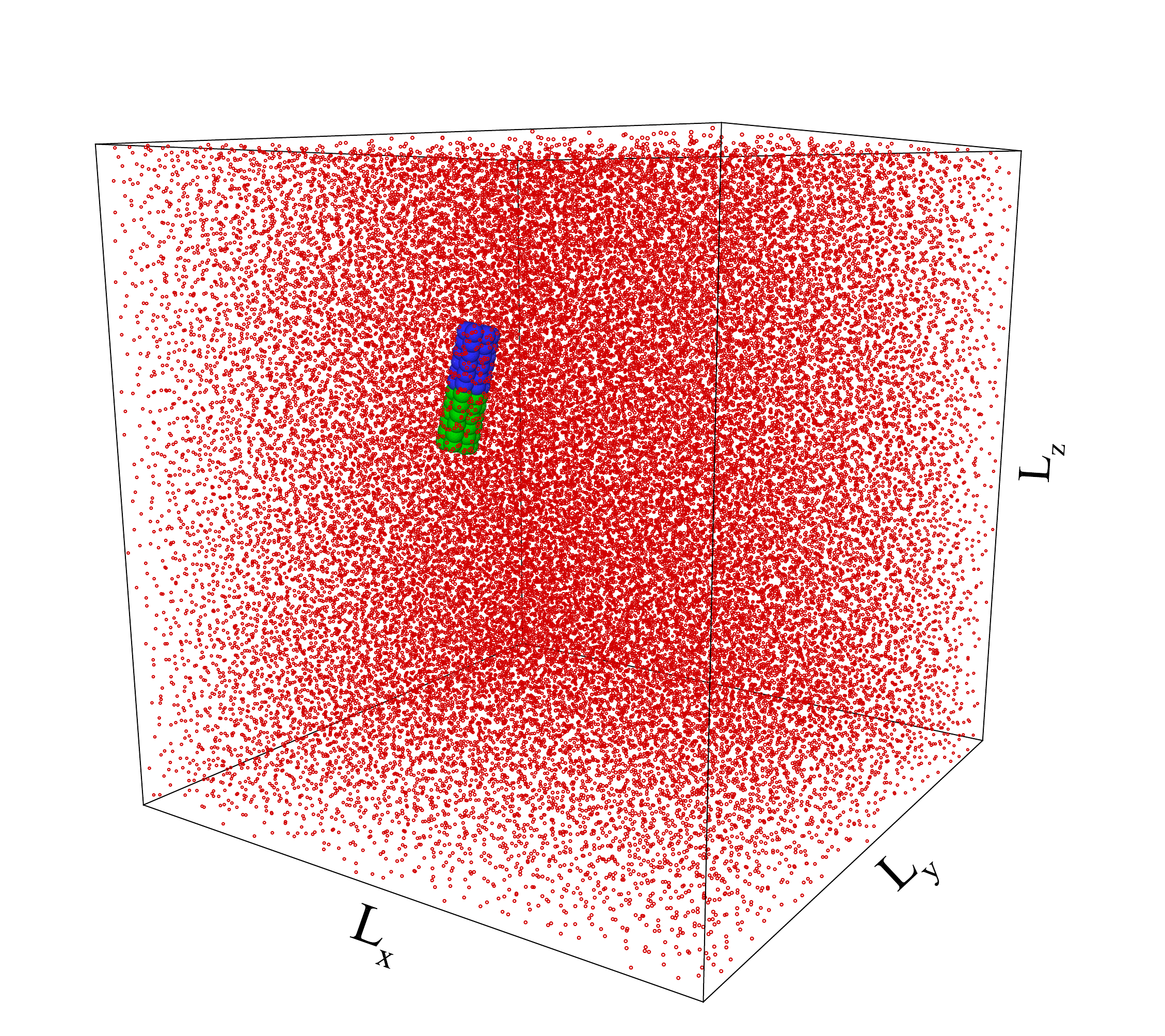}
\caption{(Color online) A snapshot of the rod-shaped Janus
nanoparticle with wetting (blue atoms) and nonwetting (green atoms)
sides and the surrounding Lennard-Jones fluid (red circles) in the
periodic box with the linear size of $39.62\,\sigma$. The number of
fluid atoms is $46\,536$. }
\label{fig:snapshot_janus_fluid}
\end{figure}

We use molecular dynamics simulations to study the translational and
rotational diffusion of a single particle in an explicit
solvent~\cite{Lammps}.   The model system consists of a rod-shaped
Janus particle immersed in a monatomic fluid at equilibrium.  A
snapshot of the system is presented in
Fig.\,\ref{fig:snapshot_janus_fluid}. In this model, any two fluid
atoms interact via the truncated Lennard-Jones (LJ) potential as
follows:
\begin{equation}
V_{LJ}(r)=4\,\varepsilon\,\Big[\Big(\frac{\sigma}{r}\Big)^{12}\!-\Big(\frac{\sigma}{r}\Big)^{6}\,\Big],
\label{Eq:LJ}
\end{equation}
where the parameters $\varepsilon$ and $\sigma$ denote the energy
and length scales of the fluid phase.   For computational
efficiency, the cutoff radius was set to $r_{c}=2.5\,\sigma$ for
fluid-fluid and fluid-solid interactions.   The fluid phase consists
of $46\,536$ monomers of mass $m$ confined in a three-dimensional
periodic cell with the linear side of $39.62\,\sigma$.    When the
finite size of a Janus particle is taken into account, the uniform
fluid density away from the particle is $\rho=0.749\,\sigma^{-3}$.
Periodic boundary conditions were applied in the $\hat{x}$,
$\hat{y}$, and $\hat{z}$ directions.   The MD simulations were
carried out in the $NVT$ ensemble, where the temperature,
$T=1.1\,\varepsilon/k_B$, was regulated by the Nos\'{e}-Hoover
thermostat with the damping time of $1.0\,\tau$. Here, $k_B$ is the
Boltzmann constant.    The equations of motion for fluid monomers
and the Janus particle were solved using the Verlet integration
algorithm~\cite{Allen87,Lammps} with a time step $\triangle
t_{MD}=0.005\,\tau$, where $\tau=\sigma\sqrt{m/\varepsilon}$ is the
characteristic LJ time.

\vskip 0.05in


The Janus rod-shaped particle was constructed by arranging $72$
atoms at vertices of 12 hexagons that are stacked together and by
adding two atoms at both ends, as shown in
Fig.\,\ref{fig:snapshot_janus}.   Hence, the particle consists of
total $74$ atoms that are fixed relative to each other and form a
symmetric rod, which moves as a rigid body in the surrounding fluid.
In this configuration, the distance between the outer hexagons along
the $\textbf{\textit{e}}_1$ axis is $6.35\,\sigma$ and the hexagon
side is $0.71\,\sigma$, which is the same as the radius of a
cylinder that contains all vertices.   The size of all particle's
atoms is the same as the size of fluid monomers.   The interaction
between solid atoms of the Janus particle and fluid monomers is also
described by the LJ potential but with different energies.  On the
wetting side, the interaction energy is fixed to
$\varepsilon_{\text{pf}}=1.0\,\varepsilon$, while on the nonwetting
side $\varepsilon_{\text{pf}}$ was varied from $0.1\,\varepsilon$ to
$1.0\,\varepsilon$.    For reference, the cases of
\textit{uniformly} wetting and nonwetting particles were also
considered, where the interaction energy with fluid monomers was set
to $\varepsilon_{\text{pf}}=1.0\,\varepsilon$ and
$\varepsilon_{\text{pf}}=0.1\,\varepsilon$, respectively.   Finally,
the total mass of the rod-shaped particle is fixed to $M=50\,m$ in
all simulations.    The particle mass $M$ was chosen to be much
larger that the mass of a fluid monomer $m$ in order to reduce
backscattering effects at short times~\cite{Levesque00,Shin10}, but,
on the other hand, this mass is small enough so that the particle
can undergo large displacements in a dense fluid, leading to an
accurate determination of diffusion coefficients without the need of
excessive computational resources.


\vskip 0.05in


The large-scale molecular dynamics simulations were performed using
the LAMMPS parallel code~\cite{Lammps}.   First, the system with the
Janus particle and the fluid was equilibrated for $2\times10^7$ MD
steps (or $10^5\,\tau$), followed by a production run of about
$10^8$ MD steps (or $0.5\times 10^6\,\tau$).   The data were
gathered in $50$ independent samples for uniform and Janus
particles. For each sample, the position of the center of mass of
the particle, its orientation vectors, as well as velocities and
positions of all particle atoms were saved every $20$ MD steps, and
these data were used later for post-processing.

\section{Results}
\label{sec:Results}

\begin{figure}[t]
\includegraphics[width=3.cm,angle=0]{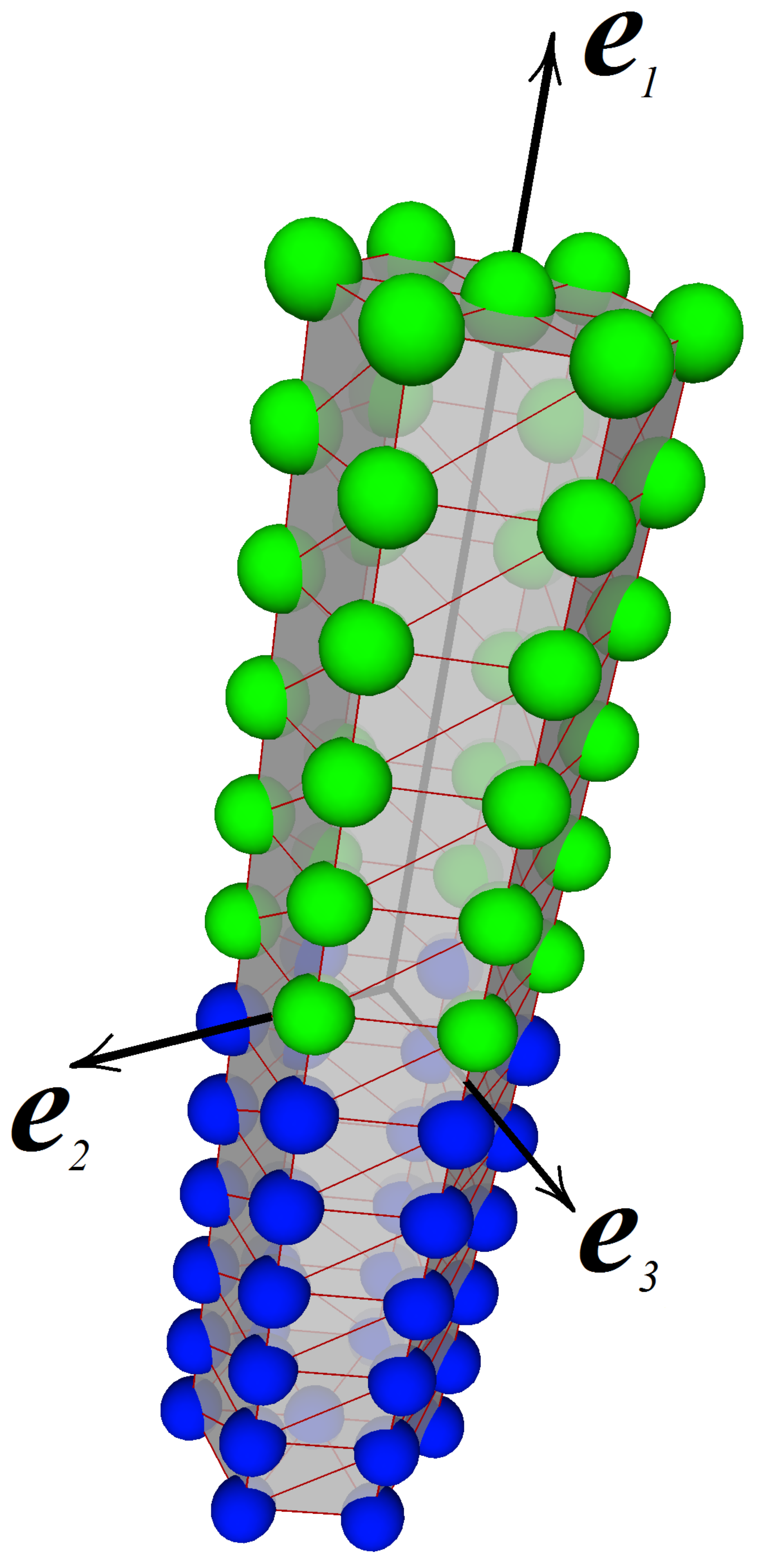}
\caption{(Color online) The Janus particle consists of $72$ atoms
fixed at the vertices of $12$ stacked hexagons and two additional
atoms at the outer faces.   The blue atoms denote the wetting side
and the green atoms indicate the nonwetting side of the rod-shaped
particle. The atoms are not drawn to scale.  The reference frame of
the particle is defined by the unit vectors $\textbf{\textit{e}}_1$,
$\textbf{\textit{e}}_2$, and $\textbf{\textit{e}}_3$. }
\label{fig:snapshot_janus}
\end{figure}
\begin{figure}[t]
\includegraphics[width=12.cm,angle=0]{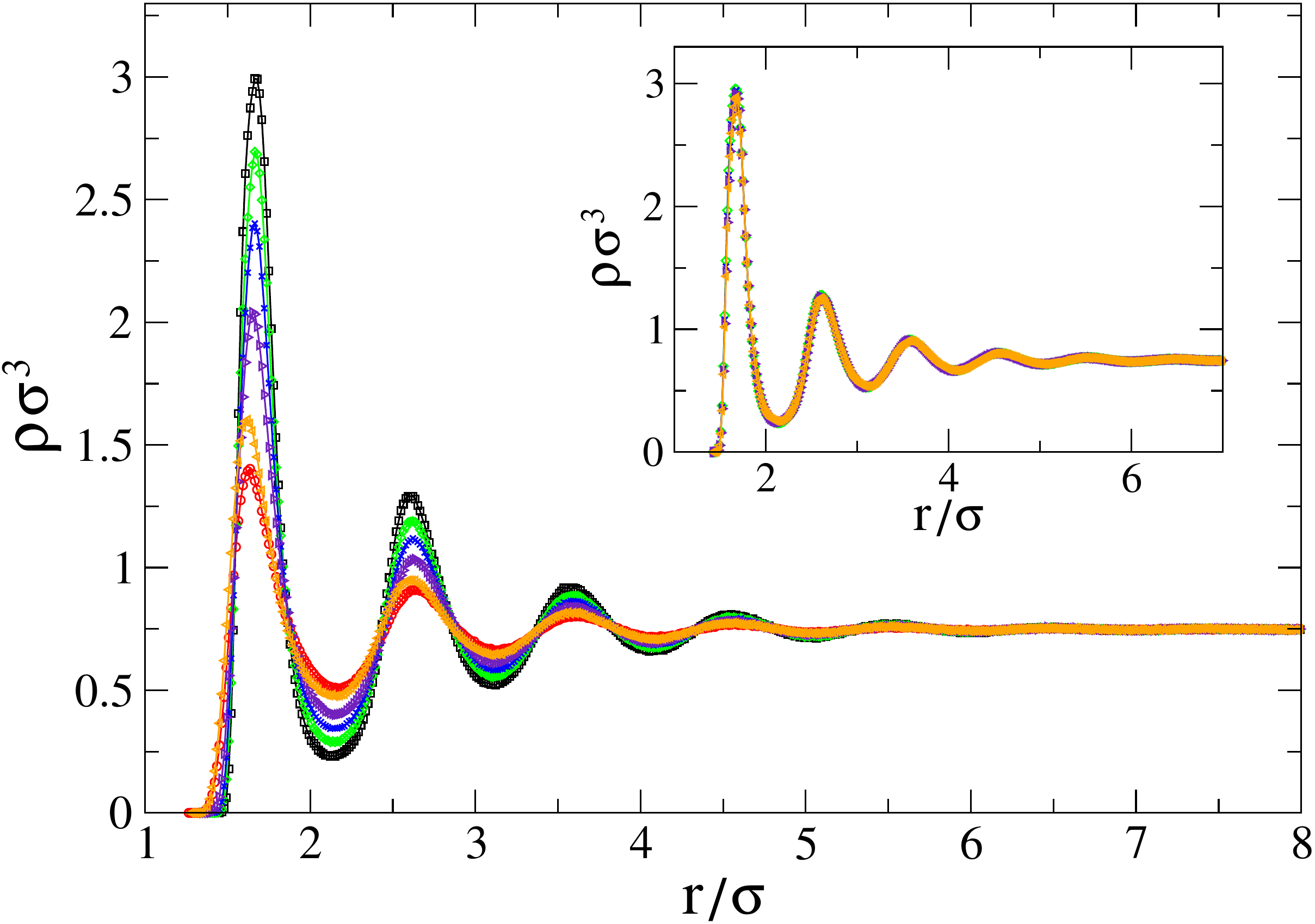}
\caption{(Color online) The radial fluid density profiles around the
uniformly nonwetting particle with $\varepsilon_{\rm
pf}=0.1\,\varepsilon$ (red), around the nonwetting side of Janus
particles with $\varepsilon_{\rm pf}=0.1\,\varepsilon$ (orange),
$\varepsilon_{\rm pf}=0.3\,\varepsilon$ (indigo), $\varepsilon_{\rm
pf}=0.5\,\varepsilon$ (blue), $\varepsilon_{\rm
pf}=0.7\,\varepsilon$ (green), and uniformly wetting particle with
$\varepsilon_{\rm pf}=1.0\,\varepsilon$ (black).    The inset shows
the radial density profiles around the wetting side of Janus
particles with the surface energy $\varepsilon_{\rm
pf}=1.0\,\varepsilon$. The color code is the same.  }
\label{fig:density_radial}
\end{figure}

We begin with a discussion of the fluid structure and local slip
boundary conditions at the particle surface.   The fluid density
profiles around the nonwetting and wetting sides of Janus particles
as well as around uniformly wetting and nonwetting particles are
presented in Fig.\,\ref{fig:density_radial}.  The density profiles
were averaged in thin cylindrical shells of radius $r/\sigma$ around
either wetting or nonwetting sides of Janus particles (see
Fig.\,\ref{fig:snapshot_janus}).   It can be seen in
Fig.\,\ref{fig:density_radial} that in all cases the density
profiles level off to the bulk value $\rho=0.749\,\sigma^{-3}$ at
$r\gtrsim 6\,\sigma$; however, a pronounced density layering is
present at smaller distances, and the amplitude of density
oscillations increases at larger surface energies.   Notice that the
height of the first peak appears to be slightly larger near the
nonwetting side of the Janus particle with $\varepsilon_{\rm
pf}=0.1\,\varepsilon$ than in the case of the uniformly nonwetting
particle with the same surface energy because fluid monomers near
the center of the Janus particle interact with atoms of the wetting
side. In other words, the contact density of the adjacent fluid
monomers varies gradually along the main axis of Janus particles due
to the finite cutoff radius of the LJ potential and difference in
surface energy at both sides.   Correspondingly, the first peak in
the density profiles around the wetting side of Janus particles is
slightly smaller for larger wettability contrast, as shown in the
inset of Fig.\,\ref{fig:density_radial}.    A similar effect of the
contact density variation was observed in the previous MD studies of
a spherical Janus particle in an explicit solvent~\cite{Kharazmi15}
and liquid films confined by surfaces of patterned
wettability~\cite{Priezjev05,Priezjev11}.

\vskip 0.05in


In order to determine the local flow boundary conditions at the
particle surface, we carried out a set of MD simulations on a
different system that consists of a monatomic fluid confined by
smooth crystalline walls (but without Janus particles).  The density
of the fluid phase and solid walls were chosen to be the same as in
the particle-fluid system described in the previous section. Special
care was taken to match the nearest-neighbor distances between
adjacent atoms at the particle surface (see
Fig.\,\ref{fig:snapshot_janus}) and the lattice constants of the
crystalline walls ($0.707\,\sigma \times 0.577\,\sigma$) that
consist of a single plane each.   The steady Poiseuille flow was
induced by applying a small force,
$f\!=\!0.0005\,\varepsilon/\sigma$, to each fluid monomer in a
direction parallel to the walls, while both walls remained at rest.
As usual, the slip length was extracted from a parabolic fit to the
velocity profile and then averaged over both
interfaces~\cite{Priezjev07,PriezjevJCP,Priezjev10}.  The results
are presented in Table\,\ref{table_ls}.  It can be seen that the
slip length increases for less wetting surfaces and it becomes
larger than the particle size.   We comment that this trend is
expected to hold for atomically smooth interfaces when the
wall-fluid interaction energy is sufficiently
large~\cite{Priezjev17}.   We also remind that in the presence of
curved surfaces, the slip boundary condition is modified by the
local radius of
curvature~\cite{Priezjev06,Niavarani10,Koplik14,Robbins16}. Finally,
the fluid viscosity was measured,
$\eta=1.66\pm0.03\,\varepsilon\tau\sigma^{-3}$, in steady flow at
density $\rho=0.749\,\sigma^{-3}$ and temperature
$T=1.1\,\varepsilon/k_B$.

\begin{table}[t]
\caption{The variation of the slip length $L_s/\sigma$ as a function
of the wall-fluid interaction energy computed at flat interfaces
between a fluid phase with the density $\rho=0.749\,\sigma^{-3}$ and
crystalline surfaces with the lattice constants $0.707\,\sigma
\times 0.577\,\sigma$ (see text for details).  The typical error
bars for the slip length are about $\pm 1.0\,\sigma$.}
 \vspace*{3mm}
 \begin{ruledtabular}
 \begin{tabular}{r r r r r r}
     \\ [-18pt]
     $\varepsilon_{\rm pf}/\varepsilon$ & $0.1$ & $0.3$ & $0.5$ & $0.7$ & $1.0~$
     \\ [5pt] \hline \\[-15pt]
     $L_s/\sigma$  &                     $36.5$ & $20.5$ & $12.0$ & $8.0$ & $4.5$
     \\ [5pt]
 \end{tabular}
 \end{ruledtabular}
 \label{table_ls}
\end{table}

\vskip 0.05in


It was previously shown that translational and rotational diffusion
coefficients for a flat-end, rigid cylinder at short lag times can
be approximated as follows:
\begin{equation}
D_{\bot}\approx\frac{k_B T}{8\pi\eta
b}\,(\text{ln}\,\omega+0.839+0.185/\omega+0.233/\omega^2),
\label{Eq:diff_perp}
\end{equation}
\begin{equation}
D_{||}\approx\frac{k_B T}{4\pi\eta
b}\,(\text{ln}\,\omega-0.207+0.980/\omega-0.133/\omega^2),
\label{Eq:diff_para}
\end{equation}
\begin{equation}
D_{r}\approx\frac{3 k_B T}{8\pi\eta
b^3}\,(\text{ln}\,\omega-0.662+0.917/\omega-0.050/\omega^2),
\label{Eq:diff_rot}
\end{equation}
where $\omega=b/a$ is the aspect ratio and parameters $a$ and $b$
are the semi-minor and semi-major axes of the cylinder, and $\eta$
is the viscosity of the solvent~\cite{Torre84,Manoharan14}.    These
predictions are expected to hold at lag times smaller than the
typical rotational relaxation time scale, since at larger lag times
the orientation of the major axis of the cylinder is decorrelated
from its initial direction and the translational diffusion becomes
isotropic.  The interpolation equations,
Eqs.\,(\ref{Eq:diff_perp})--(\ref{Eq:diff_rot}), were derived for
relatively short rigid cylinders with the aspect ratio in the range
$2\lesssim \omega \lesssim 20$~\cite{Torre84}. In our study, the
parameters $a=0.71\,\sigma$ and $b=0.32\,\sigma$ were augmented by
$0.5\,\sigma$ to take into account the finite size of LJ atoms,
which gives the aspect ratio $\omega \approx 3.04$ that was used for
the comparative analysis described below.

\vskip 0.05in

\begin{figure}[t]
\includegraphics[width=12.cm,angle=0]{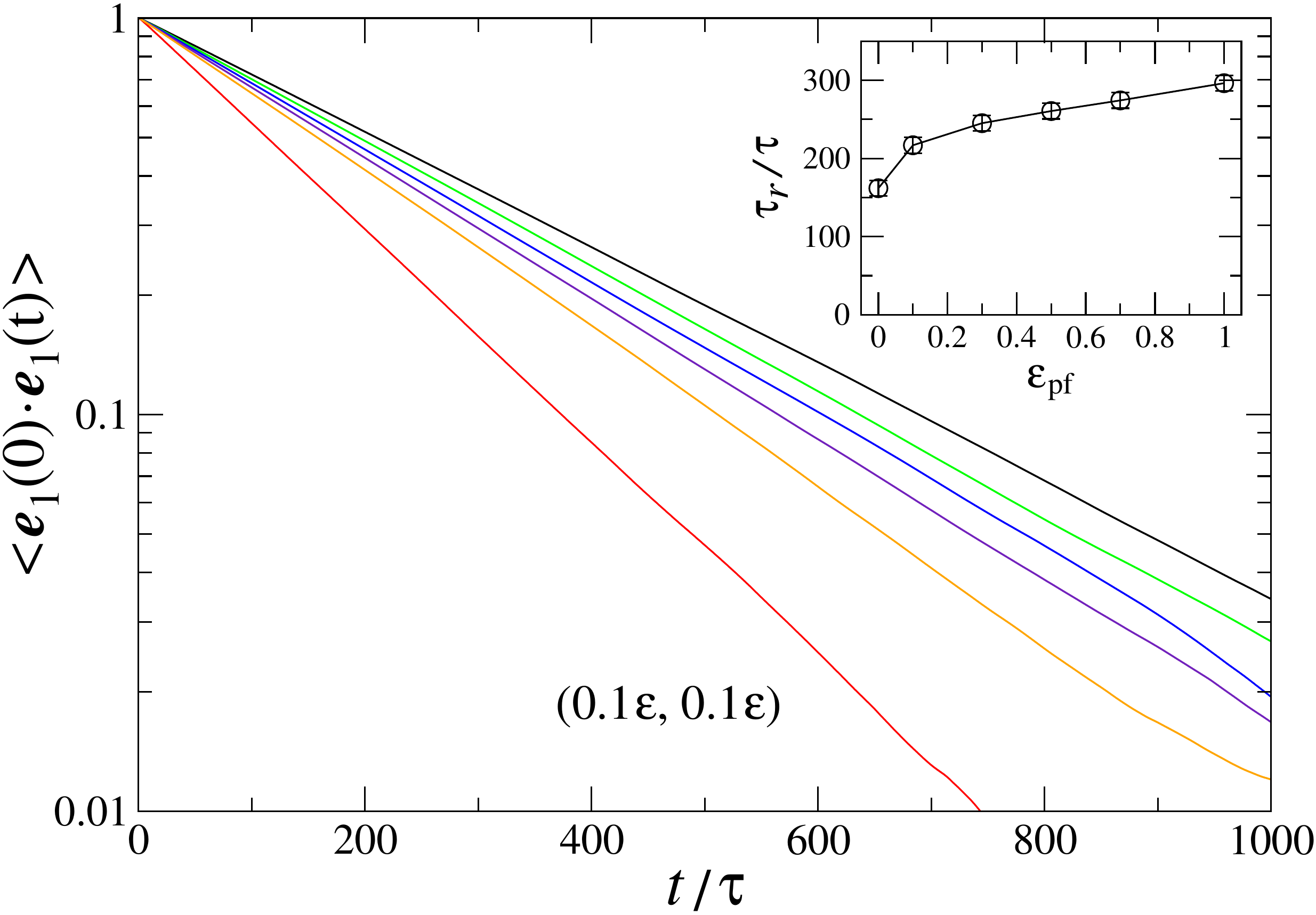}
\caption{(Color online) The time dependence of the correlation
function $\langle
\textbf{\textit{e}}_1(0)\cdot\textbf{\textit{e}}_1(t)\rangle$ for
particles with surface energies ($0.1\,\varepsilon$,
$0.1\,\varepsilon$), ($1.0\,\varepsilon$, $0.1\,\varepsilon$),
($1.0\,\varepsilon$, $0.3\,\varepsilon$), ($1.0\,\varepsilon$,
$0.5\,\varepsilon$), ($1.0\,\varepsilon$, $0.7\,\varepsilon$), and
($1.0\,\varepsilon$, $1.0\,\varepsilon$) from left to right. The
rotational relaxation time $\tau_r$ is shown in the inset as a
function of the surface energy $\varepsilon_{\text{pf}}$.  The data
point at $\varepsilon_{\text{pf}}=0$ is for a uniformly nonwetting
particle. }
\label{fig:e1e1_time}
\end{figure}

We first plot the rotational autocorrelation function of the unit
vector $\textbf{\textit{e}}_1$ along the major axis of the
rod-shaped particle $ \langle \textbf{\textit{e}}_1(0) \cdot
\textbf{\textit{e}}_1(t) \rangle$ in Fig.\,\ref{fig:e1e1_time}.   As
expected, the rotational diffusion is enhanced for the uniformly
nonwetting particle and for Janus particles with lower surface
energy at the nonwetting side.   It can be observed from
Fig.\,\ref{fig:e1e1_time} that the data for different surface
energies can be well described by the exponential decay
\begin{equation}
\langle \textbf{\textit{e}}_1(0) \cdot \textbf{\textit{e}}_1(t)
\rangle = \textit{e}^{-t/\tau_{r}},
\label{Eq:diff_e1e1_fit}
\end{equation}
where $\tau_{r}$ is the rotational relaxation time scale, and the
corresponding rotational diffusion coefficient is given by
$1/(2\tau_r)$~\cite{Manoharan14}.    The inset in
Fig.\,\ref{fig:e1e1_time} shows the variation of $\tau_{r}$ as a
function of the surface energy at the nonwetting side of the
particle. It can be seen that the rotational relaxation time
gradually varies between bounds determined by the limiting cases of
uniformly wetting and nonwetting particles.   The same data for
$\tau_{r}$ are listed in Table\,\ref{table_diff}.   We comment that
the effect of the particle shape, rod-like versus spherical, on
rotational diffusion is evident from the results reported in
Fig.\,\ref{fig:e1e1_time} in the present study and Fig.\,4 in
Ref.\,\cite{Kharazmi15}. Namely, in the case of rod-shaped
particles, the relaxation time associated with rotation of the unit
vector $\textbf{\textit{e}}_1$ is larger by a factor $\approx 1.5
\div 3$ for the same particle mass, surface wettability, fluid
density and temperature~\cite{Kharazmi15}.

\vskip 0.05in


Another way to quantify the rotational diffusion is to compute the
average mean square \textit{angular} displacement during the time
interval $t$ as follows:
\begin{equation}
\langle \Delta \vec{\varphi}^2(t) \rangle =
\frac{1}{N}\sum_{i=1}^{N}|\vec{\varphi}_i(t_0+t)-\vec{\varphi}_i(t_0)|^2
= 4 D_r t,
\label{Eq:diff_rmsd}
\end{equation}
where $D_r$ is the rotational diffusion coefficient and
$\vec{\varphi}(t)$ is the total angular displacement vector defined
by
\begin{equation}
\vec{\varphi}(t)=\int^{t}_{0}\Delta \vec{\varphi}(t^{\prime})
dt^{\prime},
\label{Eq:def_ang_disp_fi}
\end{equation}
with the magnitude of $\Delta \vec{\varphi}(t^{\prime})$ given by
$\textrm{cos}^{-1}(\textbf{\textit{e}}_1(t) \times
\textbf{\textit{e}}_1(t+t^{\prime}))$, which is the rotation angle
of the vector $\textbf{\textit{e}}_1$ during the time interval
$t^{\prime}$~\cite{Kob97,Weeks12}. In this definition, the mean
square angular displacement $\langle \Delta \vec{\varphi}^2(t)
\rangle$ is unbounded and the diffusion coefficient is evaluated in
the linear regime at large times~\cite{Kob97}.

\vskip 0.05in

\begin{figure}[t]
\includegraphics[width=12.cm,angle=0]{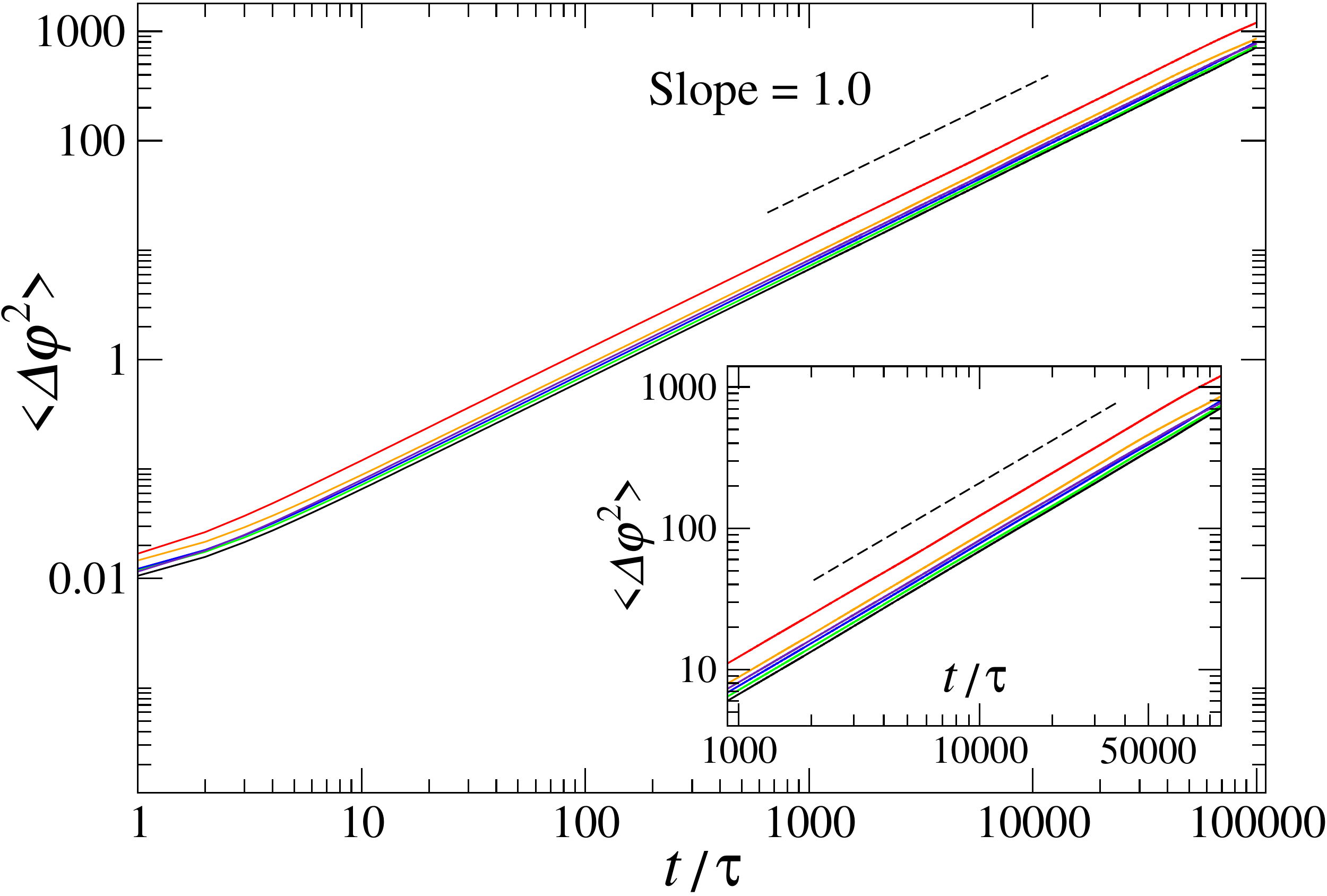}
\caption{(Color online) The mean square angular displacement
$\langle \Delta \varphi^2 \rangle$ (in units of
$\textit{rad}^{\,2}$) of the major axis (vector
$\textbf{\textit{e}}_1$) for Janus and uniform particles with
surface energies ($0.1\,\varepsilon$, $0.1\,\varepsilon$),
($1.0\,\varepsilon$, $0.1\,\varepsilon$), ($1.0\,\varepsilon$,
$0.3\,\varepsilon$), ($1.0\,\varepsilon$, $0.5\,\varepsilon$),
($1.0\,\varepsilon$, $0.7\,\varepsilon$), and ($1.0\,\varepsilon$,
$1.0\,\varepsilon$) from top to bottom.  The dashed line with slope
one is included as a reference.   The inset shows an enlarged view
of the same data at $t \geqslant 10^3\,\tau$. }
\label{fig:delfi2_time}
\end{figure}

The results for the mean square angular displacement obtained from
MD simulations are displayed in Fig.\,\ref{fig:delfi2_time} for
uniform and Janus particles.  It is clearly seen that Janus
particles with larger wettability contrast diffuse faster, and their
angular displacements are greater (smaller) than that of uniformly
wetting (nonwetting) particles.    The values of the rotational
diffusion coefficient obtained from the linear fit to
Eq.\,(\ref{Eq:diff_rmsd}) are listed in Table\,\ref{table_diff}
along with the prediction of Eq.\,(\ref{Eq:diff_rot}).   Thus, we
conclude that both methods of evaluation of the diffusion
coefficient, i.e., $D_r$ from Eq.\,(\ref{Eq:diff_rmsd}) and
$1/(2\tau_r)$ from Eq.\,(\ref{Eq:diff_e1e1_fit}) give consistent
results. However, we find that the value of the diffusion
coefficient for a uniformly wetting particle is significantly larger
than the analytical prediction of Eq.\,(\ref{Eq:diff_rot}), possibly
due to finite slip at the particle surface in MD simulations.   It
should be noted that a similar discrepancy between the prediction of
Eq.\,(\ref{Eq:diff_rot}) and MD results for the rotational diffusion
of a carbon nanotube in a LJ fluid was recently reported by Cao and
Dong~\cite{Dong14}.

\vskip 0.05in

\begin{figure}[t]
\includegraphics[width=12.cm,angle=0]{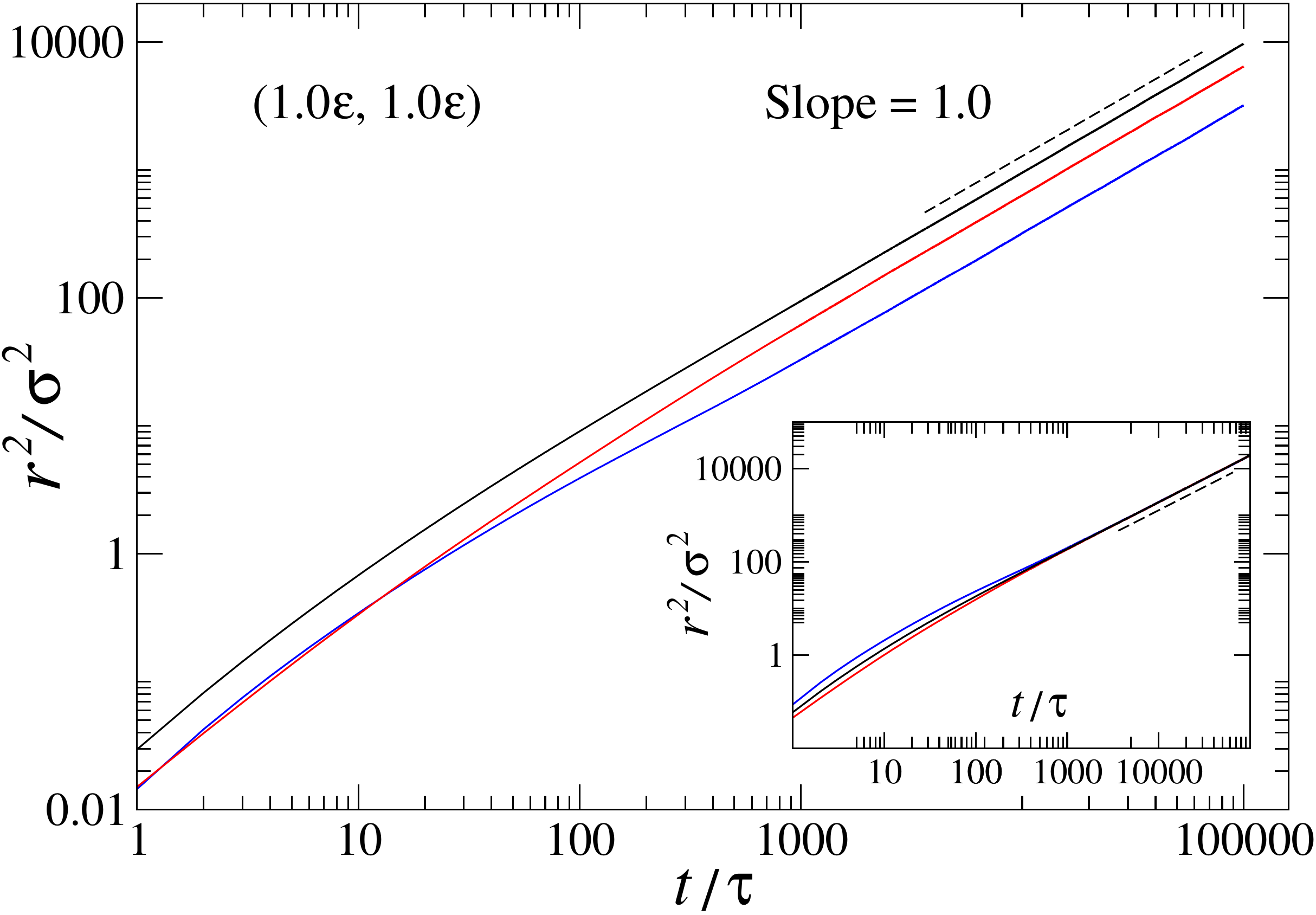}
\caption{(Color online)   The mean square displacement in the
direction parallel (blue curve) and perpendicular (red curve) to the
major axis of the uniformly wetting rod with $\varepsilon_{\rm
pf}=1.0\,\varepsilon$.   The black curve denotes the total mean
square displacement as a function of time.  The straight dashed line
with unit slope is plotted for reference.  The inset shows the same
data multiplied by different factors $6\,r^2_{||}$, $3\,r^2_{\bot}$,
and $2\,r^2_{t}$ (see text for details). }
\label{fig:msd_time_R50}
\end{figure}
\begin{figure}[t]
\includegraphics[width=12.cm,angle=0]{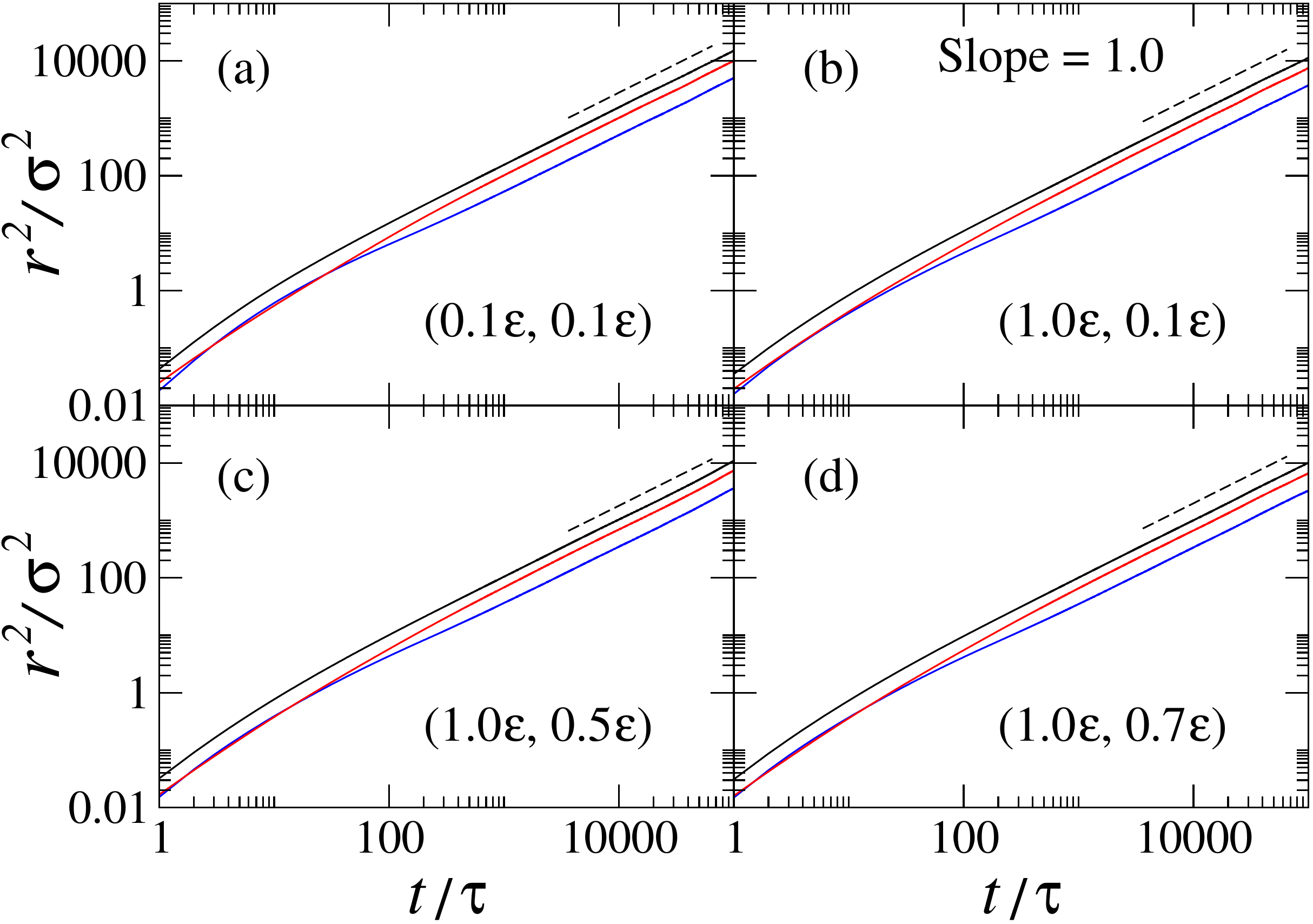}
\caption{(Color online) The averaged mean square displacements in
the direction parallel (blue curves) and perpendicular (red curves)
to the main axis of (a) uniformly nonwetting particle with
$\varepsilon_{\rm pf}=0.1\,\varepsilon$, and Janus particles with
surface energies (b) ($1.0\,\varepsilon$, $0.1\,\varepsilon$), (c)
($1.0\,\varepsilon$, $0.5\,\varepsilon$), and (d)
($1.0\,\varepsilon$, $0.7\,\varepsilon$).   The total mean square
displacement is indicated by black curves.  The dashed lines show
unit slope. }
\label{fig:msd_time_R51_Janus}
\end{figure}

The mean square displacement in the directions parallel and
perpendicular to the main axis of the uniformly wetting particle
($\varepsilon_{\rm pf}=1.0\,\varepsilon$) as well as its total mean
square displacement are presented in Fig.\,\ref{fig:msd_time_R50}.
As is evident, all curves have a unit slope at large times but with
different proportionality coefficients.   By definition, the
translational diffusion coefficient is a combination of parallel and
perpendicular diffusion coefficients, i.e.,
$D_t=(2D_{\bot}+D_{||})/3$.     To facilitate comparison, the data
for the parallel, perpendicular, and total mean square displacements
were multiplied by factors $6$, $3$, and $2$, respectively, and
replotted in the inset of Fig.\,\ref{fig:msd_time_R50}.  It can be
observed that diffusion in the direction parallel to the major axis
is faster than in the perpendicular direction at $t \lesssim
500\,\tau$, while at larger times diffusion becomes isotropic as the
orientation of the vector $\textbf{\textit{e}}_1$ is decorrelated
(see Table\,\ref{table_diff}).  As shown in
Fig.\,\ref{fig:msd_time_R51_Janus}, the same trends were observed
for uniformly nonwetting and Janus particles.   We also note that
the gradual crossover from short-time anisotropic to long-time
isotropic diffusion was observed for isolated ellipsoidal particles
in water~\cite{Lubensky06}.

\vskip 0.05in


In our analysis, the translational diffusion coefficients in the
directions parallel and perpendicular to the major axis of a
particle (along the vector $\textbf{\textit{e}}_1$) were evaluated
using the following equations:
\begin{equation}
\Delta r^2_{||}(t)=\langle (
(\textbf{r}(t_0+t)-\textbf{r}(t_0))\cdot
\textbf{\textit{e}}_1(t_0))^2 \rangle=2D_{||}t,
\label{Eq:diff_para_fit}
\end{equation}
\begin{equation}
\Delta r^2_{\bot}(t)=\langle |
(\textbf{r}(t_0+t)-\textbf{r}(t_0))\times
\textbf{\textit{e}}_1(t_0)|^2 \rangle=4D_{\bot}t,
\label{Eq:diff_perp_fit}
\end{equation}
where $t_0$ is the reference time and the brackets $\langle ..
\rangle$ indicate averaging over all $t_0$.   The
Eqs.\,(\ref{Eq:diff_para_fit})-(\ref{Eq:diff_perp_fit}) were used to
fit the data shown in Figs.\,\ref{fig:msd_time_R50} and
\ref{fig:msd_time_R51_Janus} in the range $20\,\tau \lesssim t
\lesssim \tau_r$.   The results from numerical simulations and
theoretical predictions of
Eqs.\,(\ref{Eq:diff_perp})-(\ref{Eq:diff_rot}) are reported in
Table\,\ref{table_diff}.   It can be observed that in the case of
uniformly wetting particle, the values of the diffusion coefficients
obtained from the particle trajectory agree well with predictions of
\textrm{Ref.}\,\cite{Torre84}.    We comment that this agreement
might be coincidental because of the no-slip boundary condition
assumed in derivation of
Eqs.\,(\ref{Eq:diff_perp})-(\ref{Eq:diff_rot}), while in MD
simulations the local slip length at the particle's surface is
relatively large even for the uniformly wetting case (see
Table\,\ref{table_ls}).  Generally, the data in
Table\,\ref{table_diff} follow the same trend; namely, the diffusion
coefficients are largest (smallest) for uniformly nonwetting
(wetting) particles, and the diffusion is enhanced for Janus
particles with less wetting surfaces.  The same behavior was
reported for spherical Janus and uniform particles in the previous
MD study~\cite{Kharazmi15}.


\begin{table}[t]
\caption{The diffusion coefficients for uniformly wetting and
nonwetting particles as well as Janus particles with the indicated
wettability contrast. The following parameters were evaluated from
the fit to the MD data: $D_{\bot}$ using
Eq.\,(\ref{Eq:diff_perp_fit}), $D_{||}$ using
Eq.\,(\ref{Eq:diff_para_fit}), $D_t=(2D_{\bot}+D_{||})/3$, $D_{r}$
using Eq.\,(\ref{Eq:diff_rmsd}), $\tau_r$ using
Eq.\,(\ref{Eq:diff_e1e1_fit}). The data in the last column were
computed using Eqs.\,(\ref{Eq:diff_perp})-(\ref{Eq:diff_rot}) from
\textrm{Ref.}\,\cite{Torre84}. Typical error bars are
$6\times10^{-4}\,\tau\sigma^{-2}$.  }
 \vspace*{3mm}
 \begin{ruledtabular}
 \begin{tabular}{r r r r r r r r}
   \\ [-18pt]
   $\varepsilon_{\rm pf}/\varepsilon$ & $(1.0, 1.0)$ & $(0.1, 0.1)$ & $(1.0, 0.7)$ & $(1.0, 0.5)$ & $(1.0, 0.3)$ & $(1.0, 0.1)$ & $\textrm{Ref.}\,[40]$
   \\ [5pt] \hline \\[-15pt]
   $D_{\bot}\,\tau\sigma^{-2} $ &       $0.0149$ & $0.0242$ & $0.0159$ & $0.0165$ & $0.0171$ & $0.0179$ & $0.0146$
   \\ [5pt] \hline \\[-15pt]
   $D_{||}\,\tau\sigma^{-2} $ &         $0.0177$ & $0.0291$ & $0.0186$ & $0.0197$ & $0.0203$ & $0.0209$ & $0.0174$
   \\ [5pt] \hline \\[-15pt]
   $D_{t}\,\tau\sigma^{-2}$ &           $0.0158$ & $0.0258$ & $0.0168$ & $0.0176$ & $0.0182$ & $0.0189$ & $0.0155$
   \\ [5pt] \hline \\[-15pt]
   $D_{r}\,\tau  $ &                    $0.00176$ & $0.00312$ & $0.00185$ & $0.00197$ & $0.00202$ & $0.00229$ & $0.00122$
   \\ [5pt] \hline \\[-15pt]
   $\tau_r/\tau$  &                     $296.3$ & $161.9$ & $274.3$ & $260.6$ & $245.1$ & $216.8$ & $\div~~~$
   \\ [5pt] \hline \\[-15pt]
   $\tau/2\tau_r$  &                     $0.00169$ & $0.00309$ & $0.00182$ & $0.00192$ & $0.00204$ & $0.00231$ & $\div~~~$
   \\ [5pt]
 \end{tabular}
 \end{ruledtabular}
 \label{table_diff}
\end{table}

\vskip 0.05in

\begin{figure}[t]
\includegraphics[width=12.cm,angle=0]{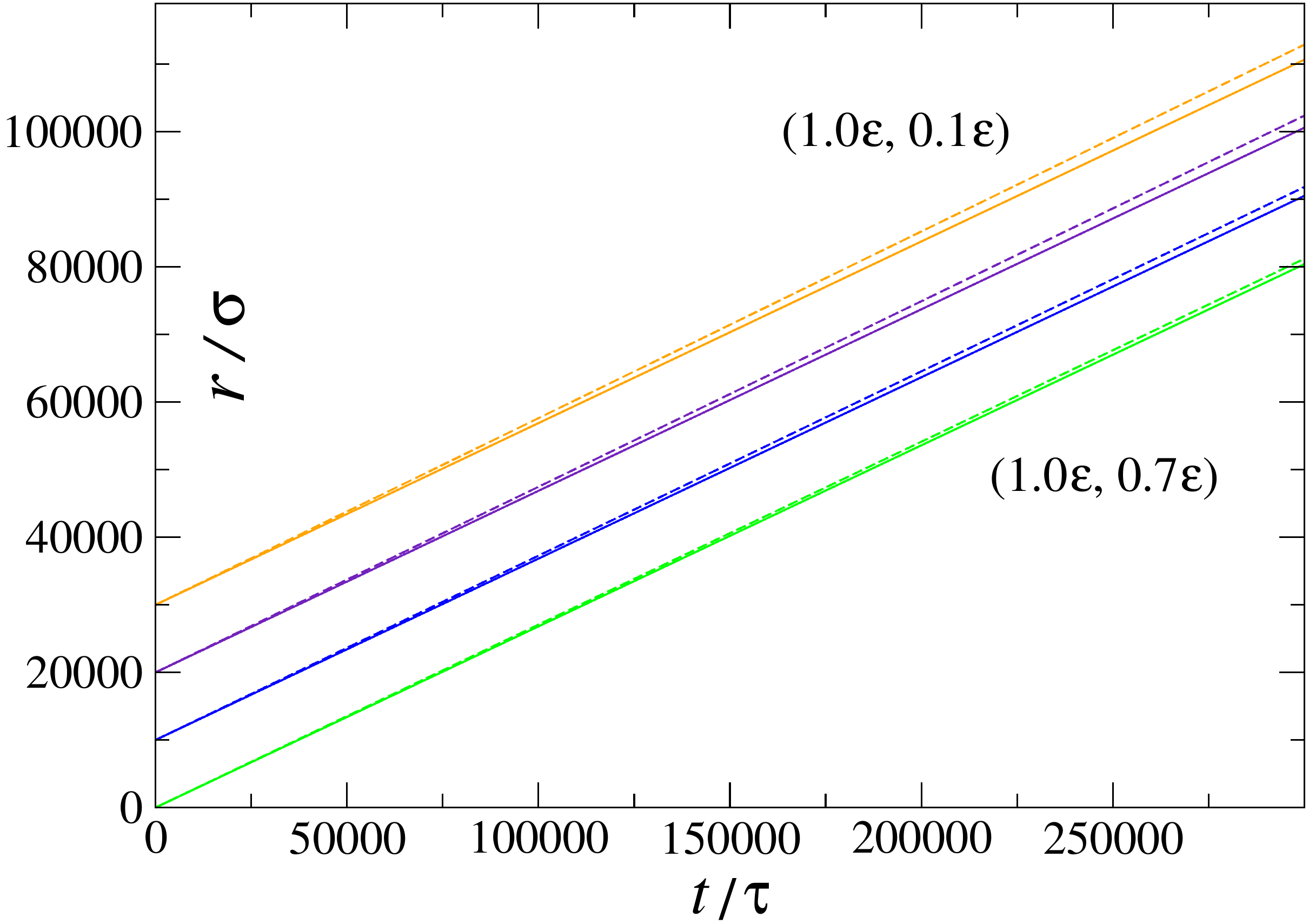}
\caption{(Color online) The total displacement, $r/\sigma$, of the
centers of nonwetting (dashed lines) and wetting (solid lines) sides
of Janus particles with surface energies ($1.0\,\varepsilon$,
$0.1\,\varepsilon$), ($1.0\,\varepsilon$, $0.3\,\varepsilon$),
($1.0\,\varepsilon$, $0.5\,\varepsilon$), ($1.0\,\varepsilon$,
$0.7\,\varepsilon$) from top to bottom.  The data for the first
three cases are displaced vertically for clarity. }
\label{fig:disp_sideAB}
\end{figure}

We next discuss more subtle aspects of the rotational dynamics
arising due to the wettability contrast and asymmetrical shape of
particles.   The time dependence of the correlation function
$\langle \textbf{\textit{e}}_1(0) \cdot \textbf{\textit{e}}_1(t)
\rangle$, shown in Fig.\,\ref{fig:e1e1_time}, provides an estimate
of the rotational relaxation time scale $\tau_r$, but it does not
describe the relative motion of the wetting and nonwetting sides.
From the analysis of particle trajectories, we computed the total
length of paths traveled by the centers of the wetting and
nonwetting sides separately.   As expected, the lengths of such
trajectories will be on average the same for uniformly wetting and
nonwetting particles (data not shown).     We also comment that the
mean square displacements of the centers of wetting and nonwetting
sides will also approach the same values at long times because they
are determined by the displacement of the particle's center of mass.
In contrast, as shown in Fig.\,\ref{fig:disp_sideAB}, the center of
the nonwetting side follows longer trajectory than the center of the
wetting side of a Janus particle.   The maximum difference in length
of the trajectories is about $2.2\times10^3\,\sigma$ at
$t=3\times10^5\,\tau$ for the Janus particle with the maximum
wettability contrast of ($1.0\,\varepsilon$, $0.1\,\varepsilon$).
These results demonstrate that during diffusive motion of Janus
particles, the nonwetting side `effectively' rotates around the
wetting side.   In other words, the center of rotation is displaced
along the main axis toward the wetting side of the Janus particle.


\vskip 0.05in

\begin{figure}[t]
\includegraphics[width=12.cm,angle=0]{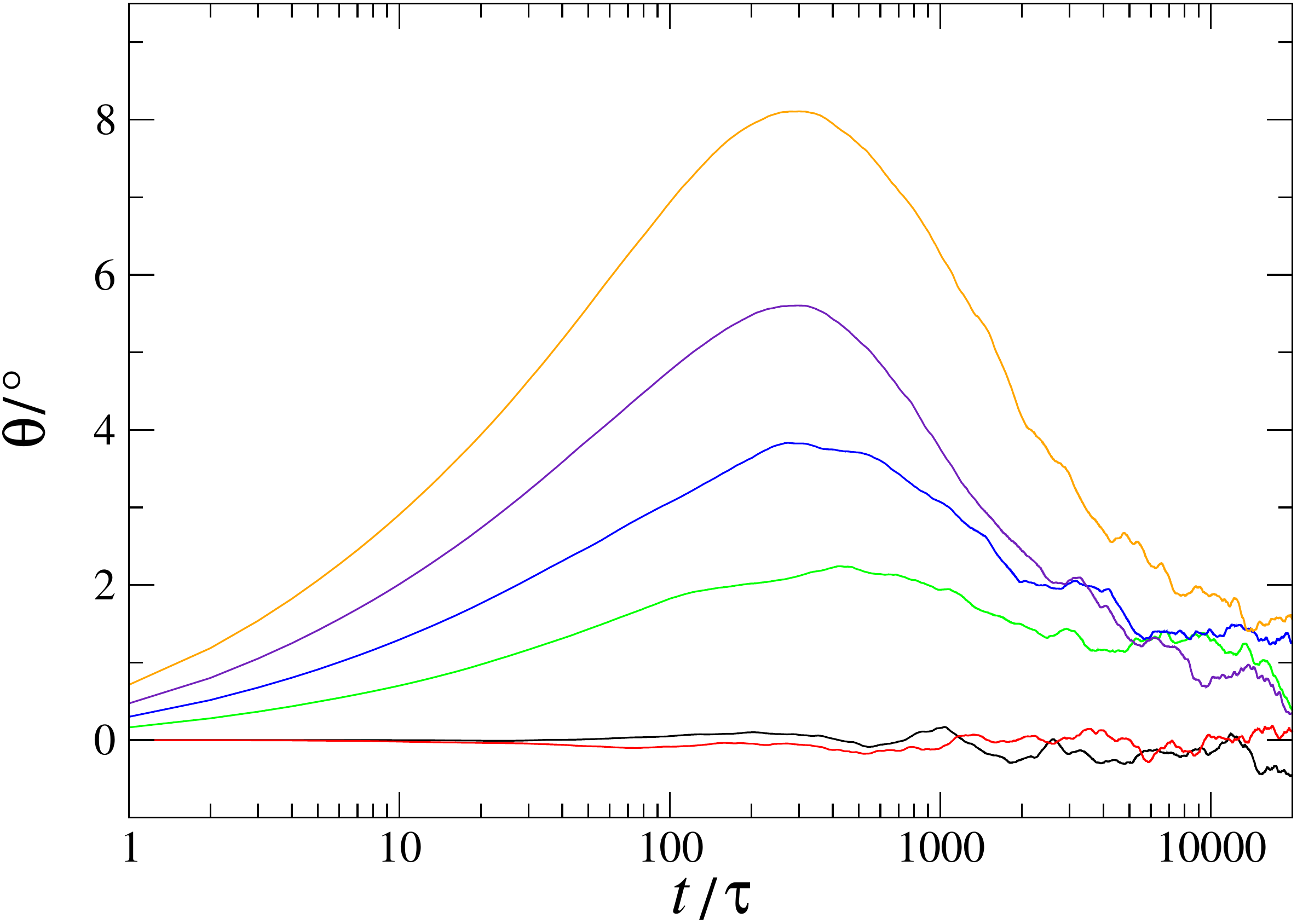}
\caption{(Color online) The angle of rotation of the vector
$\textbf{\textit{e}}_1$ along the displacement vector of the center
of mass of Janus particles with the wettability contrast
($1.0\,\varepsilon$, $0.1\,\varepsilon$), ($1.0\,\varepsilon$,
$0.3\,\varepsilon$), ($1.0\,\varepsilon$, $0.5\,\varepsilon$), and
($1.0\,\varepsilon$, $0.7\,\varepsilon$) from top to bottom. The
time dependence of the rotation angle for uniformly wetting
(nonwetting) particles is denoted by black (red) curves. }
\label{fig:angle_disp}
\end{figure}

Another peculiar feature of the diffusive motion of Janus particles
is the correlation between translation and rotation due to
asymmetric wetting~\cite{Kharazmi15}.   When, due to thermal
fluctuations, a Janus particle acquires a translational velocity in
a certain direction, its nonwetting side will tend to rotate toward
the displacement to reduce drag.   This effect can be quantified via
the average rotation angle of the major axis along the displacement
vector of the center of mass.   We first consider the displacement
vector of the center of mass $\triangle\textbf{r}$ during a time
interval $t$ and then estimate the difference between the angles
that vectors $\textbf{\textit{e}}_1(0)$ and
$\textbf{\textit{e}}_1(t)$ make with respect to
$\triangle\textbf{r}$.   In Fig.\,\ref{fig:angle_disp}, the rotation
angle averaged over particle trajectories is plotted as a function
of time for Janus and uniformly wetting particles.    It can be seen
that Janus particles indeed preferentially rotate along the
displacement, and the maximum rotation angle is attained at
intermediate times that roughly correspond to the rotational
relaxation time scales $\tau_r$ (listed in Table\,\ref{table_diff}).
Interestingly, the maximum values of the rotation angle shown in
Fig.\,\ref{fig:angle_disp} are very close to the values reported in
Fig.\,7 in Ref.\,\cite{Kharazmi15} for spherical Janus particles
with the same wettability contrast.   However, the maximum rotation
of rod-shaped Janus particles occurs at larger times than for
spherical Janus particles due to larger moment of inertia in the
former case.    Finally, as shown in Fig.\,\ref{fig:angle_disp}, the
effect of correlated rotation is absent for uniformly wetting and
nonwetting particles.

\section{Conclusions}

In this paper, molecular dynamics simulations were performed to
investigate the diffusive dynamics of Janus rod-shaped particles in
an explicit solvent. We considered the limit of low dilution where
interaction between particles can be neglected.   The Janus particle
was modeled as a rigid body where atoms are fixed at the vertices of
adjacent hexagons and form a rod with the aspect ratio of about 3.
The interaction energy between fluid monomers and particle's atoms
was set to different values on both half-sides of a Janus particle.
Two limiting cases of uniformly wetting and nonwetting particles
were also considered for reference.

\vskip 0.05in

It was shown that both rotational and translational diffusion are
enhanced for Janus particles with higher wettability contrast, while
the largest (smallest) values of the corresponding diffusion
coefficients were obtained for uniformly nonwetting (wetting)
particles. Moreover, the estimate of the diffusion coefficients for
displacement of the center of mass in the direction perpendicular
and parallel to the major axis agree well with theoretical
predictions in the case of uniformly wetting particle. The numerical
analysis of the particle trajectories revealed that the effective
center of rotation of Janus particles is displaced along the major
axis toward the wetting side.  Finally, the results of our MD
simulations indicate an unusual feature of the diffusive motion of
Janus particles; namely, the nonwetting side of the Janus particle
is rotated on average along the displacement vector of the center of
mass in order to reduce the friction force from the surrounding
fluid. Interestingly, the maximum value of the rotation angle for a
given wettability contrast of rod-shaped particles is very close to
the values reported in our previous study on diffusion of spherical
Janus particles~\cite{Kharazmi15}.

\section*{Acknowledgments}

Financial support from the National Science Foundation (CBET-1033662
and CNS-1531923) is gratefully acknowledged.   Computational work in
support of this research was performed at Michigan State
University's High Performance Computing Facility and the Ohio
Supercomputer Center.  The molecular dynamics simulations were
conducted using the LAMMPS numerical code developed at Sandia
National Laboratories~\cite{Lammps}.



\bibliographystyle{prsty}

\begin{thebibliography}{99}


\bibitem{Busseron13} E. Busseron, Y. Ruff, E. Moulin, and N. Giuseppone,
                     Supramolecular self-assemblies as functional nanomaterials,
                     Nanoscale {\bf 5}, 7098 (2013).

\bibitem{Lattuada11} M. Lattuada, T.~A. Hatton,
                     Synthesis, properties and applications of Janus nanoparticles,
                     Nano Today {\bf 6}, 286 (2011).

\bibitem{Xu12}       K. Xu, R. Guo, B. Dong, and L.-T. Yan,
                     Directed self-assembly of Janus nanorods in binary polymer mixture: towards
                     precise control of nanorod orientation relative to interface,
                     Soft Matter {\bf 8}, 9581 (2012).

\bibitem{Li13}       W. Li, B. Dong, and L.-T. Yan,
                     Janus nanorods in shearing-to-relaxing polymer blends,
                     Macromolecules {\bf 46}, 7465 (2013).

\bibitem{Pine13}     S. Sacanna, D.~J. Pine, and G.-R. Yi,
                     Engineering shape: the novel geometries of colloidal self-assembly,
                     Soft Matter {\bf 9}, 8096 (2013).


\bibitem{Rogers16}   W.~B. Rogers, W.~M. Shih, and V.~N. Manoharan,
                     Using DNA to program the self-assembly of colloidal nanoparticles and microparticles,
                     Nature Reviews Materials {\bf 1}, 16008 (2016).



\bibitem{Levesque00}  F. Ould-Kaddour and D. Levesque,
                      Molecular-dynamics investigation of tracer diffusion in a simple liquid:
                      Test of the Stokes-Einstein law,
                      Phys. Rev. E {\bf 63}, 011205 (2000).

\bibitem{Schmidt03}   J.~R. Schmidt and J.~L. Skinner,
                      Hydrodynamic boundary conditions, the Stokes-Einstein law,
                      and long-time tails in the Brownian limit,
                      J. Chem. Phys. {\bf 119}, 8062 (2003).

\bibitem{Kapral04}    S.~H. Lee and R. Kapral,
                      Friction and diffusion of a Brownian particle in a mesoscopic solvent,
                      J. Chem. Phys. {\bf 121}, 11163 (2004).

\bibitem{Schmidt04}   J.~R. Schmidt and J.~L. Skinner,
                      Brownian motion of a rough sphere and the Stokes-Einstein law,
                      J. Phys. Chem. B {\bf 108}, 6767 (2004).

\bibitem{Levesque07}  F. Ould-Kaddour and D. Levesque,
                      Diffusion of nanoparticles in dense fluids,
                      J. Chem. Phys. {\bf 127}, 154514 (2007).

\bibitem{Li09}        Z. Li,
                      Critical particle size where the Stokes-Einstein relation breaks down,
                      Phys. Rev. E {\bf 80}, 061204 (2009).

\bibitem{Shin10}      H.~K. Shin, C. Kim, P. Talkner, and E.~K. Lee,
                      Brownian motion from molecular dynamics,
                      Chem. Phys. {\bf 375}, 316 (2010).

\bibitem{Chakraborty11} D. Chakraborty,
                        Velocity autocorrelation function of a Brownian particle,
                        Eur. Phys. J. B {\bf 83}, 375 (2011).

\bibitem{Ohtori16}    Y. Ishii and N. Ohtori,
                      Molecular insights into the boundary conditions in the Stokes-Einstein relation,
                      Phys. Rev. E {\bf 93}, 050104(R) (2016).


\bibitem{Alder70}     B. Alder and T. Wainwright,
                      Decay of the velocity autocorrelation function,
                      Phys. Rev. A {\bf 1}, 18 (1970).

\bibitem{Izabela15}   K. Huang and I. Szlufarska,
                      Effect of interfaces on the nearby Brownian motion,
                      Nature Communications {\bf 6}, 8558 (2015).
\bibitem{Felderhof05} B.~U. Felderhof,
                      Effect of the wall on the velocity autocorrelation function and long-time tail of Brownian motion,
                      J. Phys. Chem. B {\bf 109}, 21406 (2005).



\bibitem{Swan08}      A.~S. Khair and J.~W. Swan,
                      On the hydrodynamics of 'slip–-stick' spheres,
                      J. Fluid Mech. {\bf 606}, 115 (2008).

\bibitem{Willmott08}  G.~R. Willmott,
                      Dynamics of a sphere with inhomogeneous slip boundary conditions in Stokes flow,
                      Phys. Rev. E {\bf 77}, 055302(R) (2008).

\bibitem{Willmott09}  G.~R. Willmott,
                      Slip-induced dynamics of patterned and Janus-like spheres in laminar flows,
                      Phys. Rev. E {\bf 79}, 066309 (2009).

\bibitem{Chan13}      Q. Sun, E. Klaseboer, B.~C. Khoo, and D.~Y.~C. Chan,
                      Stokesian dynamics of pill-shaped Janus particles with stick and slip boundary conditions,
                      Phys. Rev. E {\bf 87}, 043009 (2013).


\bibitem{Drazer15}    H. Rezvantalab, G. Drazer, and S. Shojaei-Zadeh,
                      Molecular simulation of translational and rotational diffusion
                      of Janus nanoparticles at liquid interfaces,
                      J. Chem. Phys. {\bf 142}, 014701 (2015).

\bibitem{Kharazmi15}  A. Kharazmi and N.~V. Priezjev,
                      Diffusion of a Janus nanoparticle in an explicit solvent:
                      A molecular dynamics simulation study,
                      J. Chem. Phys. {\bf 142}, 234503 (2015).

\bibitem{Zadeh16}     H. Rezvantalab and S. Shojaei-Zadeh,
                      Tilting and tumbling of Janus nanoparticles at sheared interfaces,
                      ACS Nano {\bf 10}, 5354 (2016).

\bibitem{Archereau16} A.~Y.~M. Archereau, S.~C. Hendy, G.~R. Willmott,
                      Molecular dynamics simulations of Janus particle dynamics in uniform flow,
                      arXiv:1606.02850 (2016).

\bibitem{Manoharan14} A. Wang, T.~G. Dimiduk, J. Fung, S. Razavi, I. Kretzschmar,
                      K. Chaudhary, and V.~N. Manoharan,
                      Using the discrete dipole approximation and holographic microscopy
                      to measure rotational dynamics of non-spherical colloidal particles,
                      J. Quant. Spectrosc. Radiat. Transfer {\bf 146}, 499 (2014).


\bibitem{Lammps}      S.~J. Plimpton,
                      Fast parallel algorithms for short--range molecular dynamics,
                      J. Comp. Phys. {\bf 117}, 1 (1995);
                      see also URL http://lammps.sandia.gov

\bibitem{Allen87}     M.~P. Allen and D.~J. Tildesley,
                      {\it Computer Simulation of Liquids} (Clarendon, Oxford, 1987).



\bibitem{Priezjev05}  N.~V. Priezjev, A.~A. Darhuber, and S.~M. Troian,
                      Slip behavior in liquid films on surfaces of patterned wettability:
                      Comparison between continuum and molecular dynamics simulations,
                      Phys. Rev. E {\bf 71}, 041608 (2005).

\bibitem{Priezjev11}  N.~V. Priezjev,
                      Molecular diffusion and slip boundary conditions at smooth surfaces
                      with periodic and random nanoscale textures,
                      J. Chem. Phys. {\bf 135}, 204704 (2011).



\bibitem{Priezjev07}  N.~V. Priezjev,
                      Rate-dependent slip boundary conditions for simple fluids,
                      Phys. Rev. E {\bf 75}, 051605 (2007).

\bibitem{PriezjevJCP} N.~V. Priezjev,
                      Effect of surface roughness on rate-dependent slip in simple fluids,
                      J. Chem. Phys. {\bf 127}, 144708 (2007).

\bibitem{Priezjev10}  N.~V. Priezjev,
                      Relationship between induced fluid structure and boundary slip in
                      nanoscale polymer films,
                      Phys. Rev. E {\bf 82}, 051603 (2010).

\bibitem{Priezjev17}  H. Hu, L. Bao, N.~V. Priezjev, and K. Luo,
                      Identifying two regimes of slip of simple fluids over smooth surfaces
                      with weak and strong wall-fluid interaction energies,
                      J. Chem. Phys. {\bf 146}, 034701 (2017).


\bibitem{Priezjev06}  N.~V. Priezjev and S.~M. Troian,
                      Influence of periodic wall roughness on the slip behaviour at liquid/solid
                      interfaces: molecular-scale simulations versus continuum predictions,
                      J. Fluid Mech. {\bf 554}, 25 (2006).

\bibitem{Niavarani10} A. Niavarani and N.~V. Priezjev,
                      Modeling the combined effect of surface roughness and shear rate on
                      slip flow of simple fluids,
                      Phys. Rev. E {\bf 81}, 011606 (2010).

\bibitem{Koplik14}    W. Chen, R. Zhang, and J. Koplik,
                      Velocity slip on curved surfaces,
                      Phys. Rev. E {\bf 89}, 023005 (2014).

\bibitem{Robbins16}   L. Guo, S. Chen, and M.~O. Robbins,
                      Slip boundary conditions over curved surfaces,
                      Phys. Rev. E {\bf 93}, 013105 (2016).

\bibitem{Torre84}     M.~M. Tirado, C.~L. Martinez, and J.~G. de la Torre,
                      Comparison of theories for the translational and rotational diffusion
                      coefficients of rod--like macromolecules. Application to short DNA fragments,
                      J. Chem. Phys. {\bf 81}, 2047 (1984).

\bibitem{Kob97}       S. Kammerer, W. Kob, and R. Schilling,
                      Dynamics of the rotational degrees of freedom in a supercooled liquid of diatomic molecules,
                      Phys. Rev. E {\bf 56}, 5450 (1997).

\bibitem{Weeks12}     K.~V. Edmond, M.~T. Elsesser, G.~L. Hunter, D.~J. Pine, and E.~R. Weeks,
                      Decoupling of rotational and translational diffusion in supercooled colloidal fluids,
                      PNAS {\bf 109}, 17891 (2012).


\bibitem{Dong14}      B.-Y. Cao and R.-Y. Dong,
                      Molecular dynamics calculation of rotational diffusion coefficient of a carbon nanotube in fluid,
                      J. Chem. Phys. {\bf 140}, 034703 (2014).

\bibitem{Lubensky06} Y. Han, A.~M. Alsayed, M. Nobili, J. Zhang, T.~C. Lubensky, A.~G. Yodh,
                     Brownian motion of an ellipsoid,
                     Science {\bf 314}, 626 (2006).


\end{thebibliography}

\end{document}